\documentclass[a4paper,fleqn,usenatbib]{mnras}

\usepackage{newtxtext,newtxmath}

\usepackage[T1]{fontenc}
\usepackage{ae,aecompl}
\usepackage{natbib}

\usepackage{graphicx}	
\usepackage{amsmath}	
\usepackage{amssymb}	

\title[First dredge-up and multiple populations]{Photometric 
characterization of multiple populations in star clusters: The impact of the first dredge-up}

\author[M. Salaris et al.]{Maurizio Salaris,$^{1}$\thanks{E-mail: M.Salaris@ljmu.ac.uk}
Chris Usher,$^{1}$
Silvia Martocchia,$^{1,2}$
Emanuele Dalessandro,$^{3}$
\newauthor
Nate Bastian,$^{1}$
Sara Saracino,$^{1}$
Santi Cassisi,$^{4,5}$
Ivan Cabrera-Ziri,$^{6}$\thanks{Hubble fellow}
and Carmela Lardo$^{7}$
\\
$^{1}$Astrophysics Research Institute, Liverpool John Moores University, 146 Brownlow Hill, Liverpool L3 5RF, UK\\
$^{2}$European Southern Observatory, Karl-Schwarzschild-Strasse 2, D-85748 Garching bei M\"unchen, Germany\\
$^{3}$INAF-Osservatorio di Astrofisica e Scienza dello Spazio di Bologna, Via Gobetti 93/3, Bologna, II-40129, Italy\\
$^{4}$INAF-Osservatorio Astronomico d'Abruzzo, via M. Maggini, sn. 64100, Teramo, Italy\\
$^{5}$INFN - Sezione di Pisa, Largo Pontecorvo 3, 56127 Pisa, Italy\\
$^{6}$Harvard-Smithsonian Center for Astrophysics, 60 Garden Street, Cambridge, MA 02138, USA\\
$^{7}$Laboratoire d'astrophysique, Ecole Polytechnique F\'ed\'erale de Lausanne (EPFL),
Observatoire de Sauverny, CH-1290 Versoix, Switzerland
}

\date{Accepted XXX. Received YYY; in original form ZZZ}

\pubyear{2015}

\begin{document}
\label{firstpage}
\pagerange{\pageref{firstpage}--\pageref{lastpage}}
\maketitle

\begin{abstract}
  The existence of 
  star-to-star light-element abundance variations (multiple populations, MPs)
  in massive Galactic and extragalactic star clusters older than about 2~Gyr is
  by now well established. Photometry of red giant branch (RGB) stars has been and still is  
  instrumental in enabling the detection and characterization of cluster MPs,  
  through the appropriate choices of filters, colours and colour combinations, that are mainly
  sensitive to N and --to a lesser degree-- C stellar surface abundances.
  An important issue not yet properly addressed is that the translation of the observed
  widths of the cluster RGBs to abundance spreads must account for the effect of the first dredge-up
  on the surface chemical patterns, hence on the spectral energy distributions of stars belonging to the various MPs. 
  We have filled this gap by studying theoretically the impact of the dredge-up on the predicted widths of RGBs 
  in clusters hosting MPs. We find that for a given initial range of N abundances, the first dredge up reduces the predicted
  RGB widths in N-sensitive filters
  compared to the case when its effect on the stellar spectral energy distributions is not accounted for.
  This reduction is a strong function of age and has also a dependence on metallicity. The net effect is an underestimate
  of the initial N-abundance ranges from RGB photometry if the first dredge-up is not accounted for in the modelling,
  and also the potential determination of spurious trends of N-abundance spreads with age. 
\end{abstract}




\section{Introduction}

During the last 10-15 years both spectroscopic and photometric observations have definitely established
that individual Galactic globular clusters (GCs) host multiple populations (MPs) of stars, characterised
by C-N, O-Na (and also, but not always, Mg-Al) anticorrelations and He abundance spreads 
\citep[see, e.g.][]{gcb, m17, He18, bl18, g19}. 
Scenarios for the origin of MPs 
\citep[reviewed, e.g, by][]{r15, bl18} generally invoke more than one episode of star formation, envisaging that 
stars with CNONa (and He) abundance ratios similar
to those observed in halo field stars are the first objects to form (P1 stars), whilst
stars enriched in N and Na (and He) and depleted in C and O formed later (P2 stars). These P2 stars are supposed to form out of  
chemically processed material ejected by some class of massive P1 stars, usually denoted as polluters.
To date, none of the proposed polluters can explain quantitatively the full ensemble of chemical 
patterns observed in individual GCs \citep[][]{r15, bl18}.

Photometric \citep[see, e.g.,][and references therein]{larsen14, dale16, gilligan19, lagioia, nardiello19}
and to a lesser extent spectroscopic \citep{m09} observations  
have also shown that MPs are not confined only to Galactic GCs, but are present also in  
old clusters of the Magellanic Clouds, Fornax and M31.
Integrated spectroscopy of old extragalactic clusters in M31 also confirms the signature of MPs 
amongst old massive stellar clusters \citep[see, e.g.,][]{s13}

A further recent development has seen the detection of MPs in resolved massive extragalactic clusters 
down to ages of $\sim$2~Gyr through spectroscopy \citep{hollyhead17, h19}, and to a much larger extent
photometry \citep[see, e.g.,][and references therein]{n17, n17b, martocchia,lagioia,nardiello19,mart19,milone19}.
Even more recently \citet{bastian19} found a signature of MPs also in an intermediate-age massive clusters
belonging the galaxy NGC~1316, from integrated spectroscopy.
These new results are extremely important, for they clearly point to a connection between the formation of old GCs and much 
younger massive clusters.

As mentioned above, photometry plays a crucial role in the detection and characterization of MPs
in resolved clusters. 
Several authors have shown, both empirically and theoretically \citep[see, e.g.,][]{sw06, marino, yong, sbordone, cmp13, dale16, 
mucc16, m17, dale18, jwst},
that appropriate choices of filters, colours and
colour combinations (these latter usually denoted as \textit{pseudocolours}) are sensitive
to the abundance of mainly nitrogen (plus carbon and oxygen to a much lesser extent) in the atmospheres of the
target stars, and can clearly detect the presence of MPs
\citep[see, e.g.,][]{sumo, p15, m17, n17, jwst}. Due to the distance of the targets, red giant branch (RGB) stars are typically 
observed for photometric (and also spectroscopic) MP detection. 

By analyzing results for a number of Magellanic Cloud (MC) clusters covering a large range of ages (2-8~Gyr),
\citet{martocchia} and \citet{mart19} -- hereinafter M19-- found in their
sample a general trend between the measured width of a cluster RGB 
and the cluster age. They considered the pseudocolours $C_{F343N,F438W,F814W} \equiv (F343N-F438W)-(F438W-F814W)$ and
$C_{F336W, F438W, F343N} \equiv (F336W-F438W)-(F438W-F343N)$ in the filter systems of the WFC3 and ACS (for $F814W$)
cameras on board the {\sl Hubble Space Telescope} (HST) --both sensitive to the abundance of N in the stellar spectra--
and found that the RGB width shows a general increase with increasing age in their cluster sample.
A natural explanation for this occurrence is that the N spread in these massive clusters increases with increasing
cluster age.

However, there is an important additional phenomenon to consider when translating the observed RGB widths to
N abundance spreads,
that has been so far largely unexplored. 
The samples of cluster stars considered in M19 are distributed between the base of the
RGB and approximately the RGB bump. This range covers almost the entire evolution through the 
first dredge-up \citep[FDU -- see, e.g.,][and references therein]{k14, s15}, that starts on the subgiant branch
and ends below the RGB bump level.
During the FDU the surface N abundance increases compared to the initial value, and from basic
stellar physics we expect 
this increase to depend on the initial nitrogen abundances. The variation of the surface abundances due to the FDU
impacts the stellar spectral energy distributions, hence the predicted colours and pseudocolours sensitive
to this element.
The upshot is that the observed RGB widths in M19 are determined by a combination of the
initial N spreads plus the effect of the dredge up.
This is true for any colour or pseudocolour sensitive to the surface N abundance in RGB stars, like 
$C_{F275W, F336W, F438W} \equiv (F275W-F336W)-(F336W-F438W)$ \citep[see, e.g.,][]{m17, sara19}, or 
$C_{F336W,F343N,F438W} \equiv(F336W-F343N)-(F343N-F438W)$ \citep{zenn19}, and
$C_{F275W,F343N,F438W} \equiv(F275W-F343N)-(F343N-F438W)$ \citep{milone19}, devised to detect
MPs in GCs and younger MC massive clusters by means of the so-called \lq{chromosome maps\rq}.

%
%

In this paper we explore this issue showing qualitatively, and in case of M19 results also
quantitatively, the important role played by the FDU when translating to N abundance spreads
the observed RGB widths in magnitude-pseudocolour diagrams and chromosome maps. 
Section~\ref{models} presents our theoretical calculations and the effect of the FDU
on the surface abundances of MP stars, followed in Sect.~\ref{analysis}
by an analysis of the impact of the FDU
on the $C_{F275W, F336W, F438W}$, $C_{F336W, F343N, F438W}$ and $C_{F275W, F343N, F438W}$ pseudocolours used in
the chromosome maps to disentangle cluster MPs, plus a quantitative estimate of the effect on
the $C_{F343N,F438W,F814W}$ pseudocolour employed by M19.
Conclusions follow in Sect.~\ref{conclusions}.

\section{Models}
\label{models} 

We have employed the BaSTI  isochrones \citep{basti} for two P1 solar scaled chemical compositions with 
[Fe/H]=$-$0.35 and [Fe/H]=$-$1.3, respectively, representative of the MC clusters studied by M19.
We have also isochrones with the same stellar evolution code but
for a reference N-enhanced P2 compositions with [C/Fe] =$-$0.28 [N/Fe] = +0.80 [O/Fe] = $-$0.28
(that keeps the CNO sum unchanged compared to
the P1 metal distribution) and the same two [Fe/H] values.
As discussed by \citet{s15} and \citet{bm17}, BaSTI models reproduce
the general observed trends of post-FDU [C/N] ratios
as a function of age, measured in field halo stars and a sample of open clusters of various ages.
%
%

Using the model RGB effective temperatures, surface gravity and surface chemical abundances as inputs, we have calculated
model atmospheres and synthetic spectral energy distributions as in \citet{mart17}, employing
the ATLAS12 \citep{sao70, k05} and SYNTHE \citep{sao79, sao81} codes, respectively. From these spectral energy distributions we have then computed 
bolometric corrections to transform the bolometric luminosities of the stellar models to magnitudes in the
HST WFC3 and ACS filter systems.

All calculations show that at 
the end of the main sequence the outer convection zone engulfs increasingly deeper regions,
dredging to the surface matter partially processed by H-burning in the core  --the FDU.
In this phase the N abundance slowly increases (and the carbon abundance slowly decreases, but to a much smaller extent), until 
the maximum penetration of the convective envelope is reached. At this point the
inner convective boundary starts to retreat towards the surface, leaving behind a chemical discontinuity. This discontinuity is eventually
crossed by the advancing (in mass) H-burning shell, causing the appearance of the RGB
bump in the luminosity function of old stellar populations
\citep[see, e.g.,][]{cs13}. 
The change of surface nitrogen abundance
(and carbon, whilst the oxygen abundance is essentially never altered in the age range investigated here)
during the FDU depends on the mass of the star \citep[hence the population age,
  see e.g.][and references therein]{s15}, but also on the initial abundance pattern.
The reason is that during the FDU the convective envelope reaches layers
where the abundances of C and N attained the equilibrium values of the CN cycle  
during the main sequence. The equilibrium abundance of N is typically higher (and the C abundance lower) than the
standard solar scaled (or $\alpha$-enhanced) counterpart for a given total metallicity, hence the FDU causes an increase of
surface N (and decrease of C). 
When the initial metal mixture is N-enhanced (and carbon depleted) the equilibrium abundance of
N (and C) becomes more comparable to the initial one, and the effect
of the FDU is much less appreciable or negligible.

\begin{figure}
	\includegraphics[width=\columnwidth]{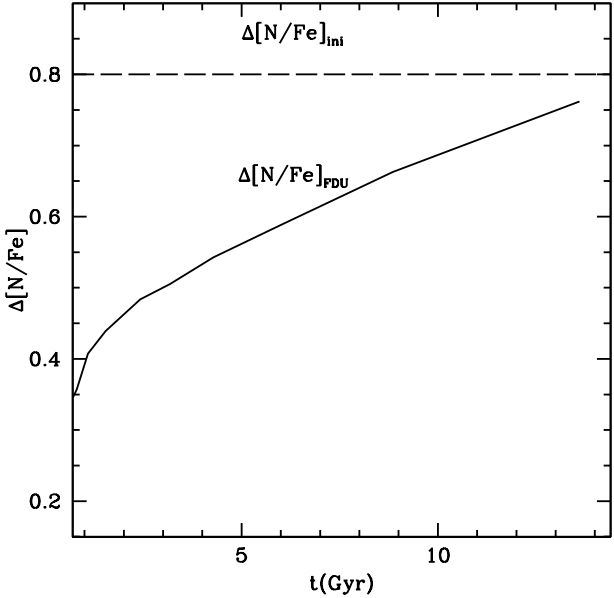}
        \caption{Initial value ($\Delta {\rm [N/Fe]_{ini}}$=0.8 -- dashed line) of the difference in the surface [N/Fe]
          ratio for a set of bimodal
          populations with [Fe/H]=$-$1.3 and various ages t,
          and the corresponding difference $\Delta {\rm [N/Fe]_{FDU}}$ in the surface abundances after the FDU
          (solid line -- see text for details).}
    \label{figDN}
\end{figure}


Figure~\ref{figDN} shows the run with age (ages between 1.0 and 13.5~Gyr) of $\Delta$[N/Fe], defined as
the difference of surface [N/Fe] between a population with with N-enhanced P2 composition
and a coeval one with P1 composition
([Fe/H]=$-$1.3 in this example) as inferred from our models.
We show both the initial value $\Delta {\rm [N/Fe]_{ini}}$ (the same for all ages) and the corresponding 
surface abundance differences at the completion of the FDU ($\Delta {\rm [N/Fe]_{FDU}}$).

 
The values of $\Delta {\rm [N/Fe]_{FDU}}$ are lower than $\Delta {\rm [N/Fe]_{ini}}$,
and display a clear trend with age, despite the fact that $\Delta {\rm [N/Fe]_{ini}}$ is the same at all ages.
For the younger populations $\Delta {\rm [N/Fe]_{FDU}}$ is much smaller than $\Delta {\rm [N/Fe]_{ini}}$,
but with increasing age it comes progressively closer to its initial value.
The reason is that in RGB models with P1 initial N abundance, the surface [N/Fe] 
at the end of the FDU increases with decreasing age 
\citep[see, e.g.][]{s15}, whilst the impact of the FDU is much smaller or negligible in N-enhanced populations.
The effect of the FDU on the surface carbon abundance is also small or negligible in P2 models with initial C-depleted abundances, whilst
in P1 models the surface [C/Fe] at the end of the FDU gets progressively lower with decreasing age.

Summarizing, for bimodal coeval populations with fixed $\Delta {\rm [N/Fe]_{ini}}$ (independent of age) and ages between 1.0 and 13.5~Gyr,
the difference $\Delta {\rm [N/Fe]_{FDU}}$ measured at the end of the FDU is predicted to be lower than the initial value,
showing a trend with age -- $\Delta {\rm [N/Fe]_{FDU}}$ decreasing for decreasing age. This general behaviour 
is true irrespective of the exact value of [Fe/H].

\section{Analysis}
\label{analysis}

We start here discussing the impact of the FDU on the representative N-sensitive 
$C_{F275W, F336W, F438W}$, $C_{F336W, F343N, F438W}$ and $C_{F275W, F343N, F438W}$ pseudocolours used in the chromosome maps to detect MPs from 
cluster photometry. In a chromosome map the total width of the cluster RGB in one of these pseudocolors is normalized to
the value taken two magnitudes above the main sequence
turnoff in the $F814W$ filter \citep{m17}. This is a level where typically the FDU has either
already started or is essentially completed, but still below the RGB
bump, beyond which extra mixing processes that further affect the surface C and N abundances 
appear to be efficient, at least in low mass stars \citep[see, e.g.,][and references therein]{lagarde19}.

\begin{figure}
	\includegraphics[width=\columnwidth]{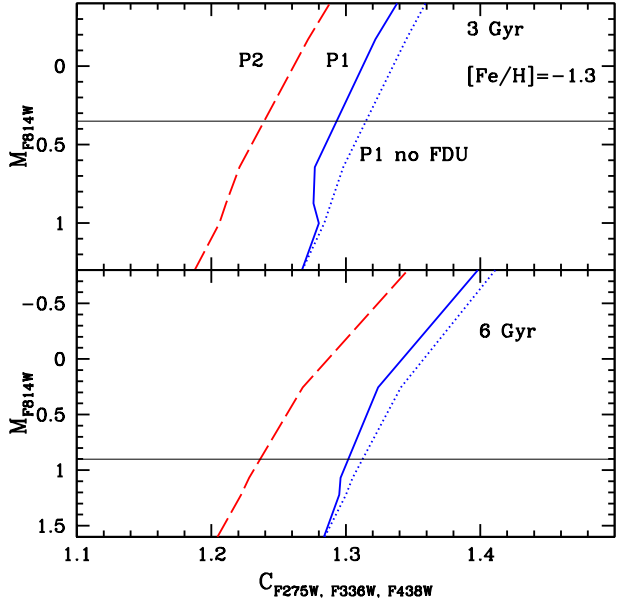}
        \caption{P1 (solid lines) and P2 (dashed lines) RGBs (at [Fe/H]=$-$1.30) in the $M_{F814W}-C_{F275W, F336W, F438W}$ diagram
          for ages equal to 3 (upper panel) and 6 (lower panel) Gyr,
          $\Delta {\rm [N/Fe]_{ini}}$=0.8, accounting for the effect of FDU on the spectral energy distributions.
          The dotted lines display P1 RGBs calculated without considering the variation of N (and C) due to the FDU.
          The horizontal thin lines mark the level corresponding to two magnitudes above the main sequence
          turnoff, where the width of the RGB is taken in the chromosome maps (see text for details).}
    \label{figchro}
\end{figure}

\begin{figure}
	\includegraphics[width=\columnwidth]{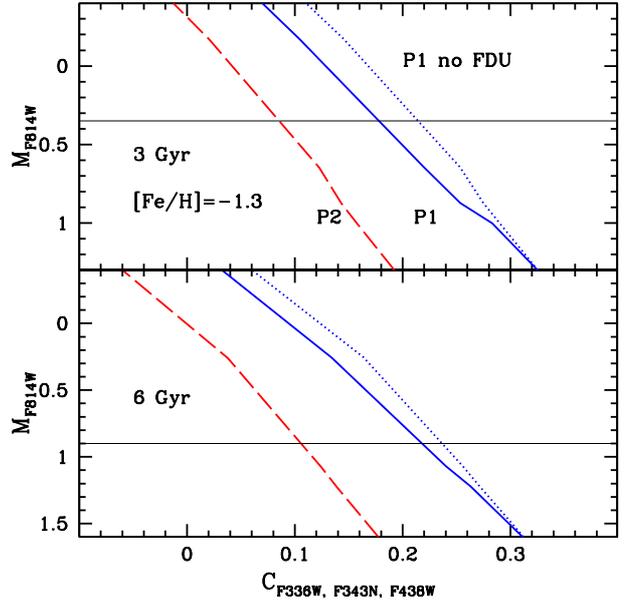}
        \caption{As Fig.~\ref{figchro} but for the $M_{F814W}-C_{F336W, F343N, F438W}$ diagram.}
    \label{figchrob}
\end{figure}

\begin{figure}
	\includegraphics[width=\columnwidth]{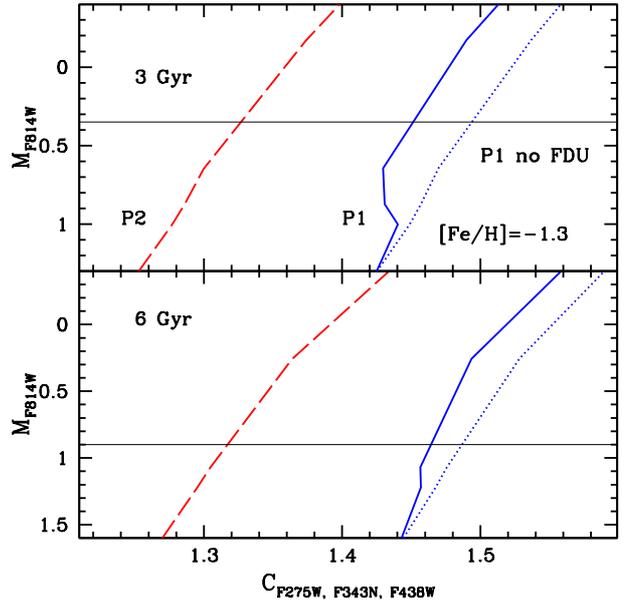}
        \caption{As Fig.~\ref{figchro} but for the $M_{F814W}-C_{F275W, F343N, F438W}$ diagram.}
    \label{figchromilone}
\end{figure}

As long as cluster ages are of the order of 10-13~Gyr, the effect of the FDU on the surface abundances --hence the
RGB $C_{F275W, F336W, F438W}$, $C_{F336W, F343N, F438W}$ and $C_{F275W, F343N, F438W}$ values-- is basically negligible, especially at low metallicity.
The situation is however different for intermediate-age clusters, as shown in 
Figs.~\ref{figchro}, ~\ref{figchrob} and ~\ref{figchromilone} that display $M_{F814W}$-$C_{F275W, F336W, F438W}$, 
$M_{F814W}-C_{F336W, F343N, F438W}$ and $M_{F814W}-C_{F275W, F343N, F438W}$ diagrams 
for two P1-P2 bimodal populations (at a representative [Fe/H]=$-$1.3) with ages of 3 and 6 Gyr, respectively, including also P1 RGBs calculated
without accounting for the FDU in the spectral energy distributions.
In the case of the P2 models the FDU never changes appreciably the surface abundances.  
The level at which the total RGB width is normalized in the chromosome maps to compare different clusters 
(two magnitudes above the main sequence turn off in the $F814W$ filter) is also marked.

At the beginning of the FDU the pseudocolour difference between P1 and P2 RGBs 
is essentially the same for these two ages, in all three diagrams; when  
moving to brighter magnitudes the FDU progresses, causing a reduced 
[N/Fe] (and [C/Fe]) difference between P1 and P2 stars at a given brightness, and a decreased separation of the sequences.
The effect is stronger at younger ages, because of the increasing impact of the FDU on the surface abundances (see Fig.~\ref{figDN}).
Clearly, any interpretation in terms of $\Delta {\rm [N/Fe]_{ini}}$ of the RGB widths in the chromosome maps  
must account for the effect of the FDU on the surface [N/Fe] and pseudocolours, for ages lower than typical GC ages.

\begin{figure}
	\includegraphics[width=\columnwidth]{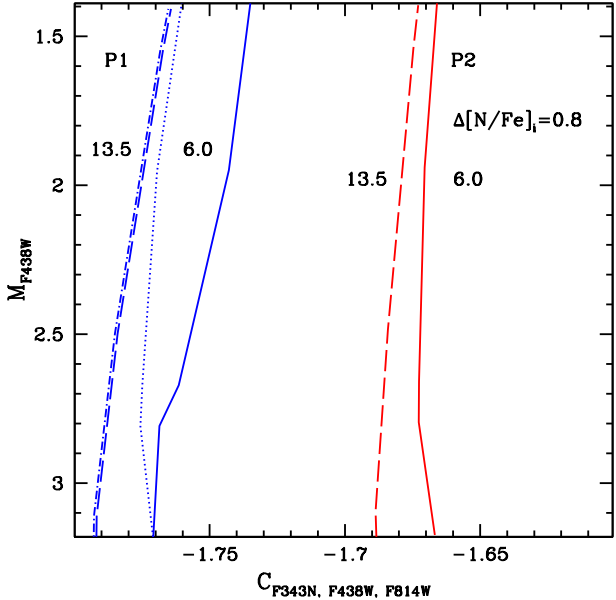}
        \caption{P1 and P2 RGBs (at [Fe/H]=$-$1.30) in the $M_{F438W}-C_{F343N,F438W,F814W}$ diagram
          for 6 (solid lines) and 13.5 (dashed lines) Gyr,
          $\Delta {\rm [N/Fe]_{ini}}$=0.8, accounting for the FDU (that has a negligible effect on P2 RGBs).
          The dotted and dash-dotted blue lines display
          P1 RGBs calculated without considering the variation of N (and C) due to the FDU, for 6 and 13.5 Gyr, respectively 
          (see text for details).}
    \label{figCMD}
\end{figure}

To give at least one quantitative example of the role played by the FDU when inferring and/or
comparing $\Delta {\rm [N/Fe]_{ini}}$ values amongst different clusters using N-sensitive photometric properties, 
we consider the $C_{F343N,F438W,F814W}$ pseudocolour employed by M19.
Figure~\ref{figCMD} shows theoretical 
RGBs for two bimodal P1-P2 populations with  ages equal to 6 and 13.5 Gyr at a representative [Fe/H]=$-$1.3, 
and $\Delta {\rm [N/Fe]_{ini}}$=0.8, plotted in the $M_{F438W}-C_{F343N,F438W,F814W}$ diagram used by M19.
The displayed range of $M_{F438W}$ magnitudes corresponds approximately to the range
employed in M19 analysis, that roughly encompasses the entire development of the FDU.

As for the cases discussed before, at a given age the $C_{F343N,F438W,F814W}$ separation between P1 and P2 RGBs decreases with
decreasing magnitude, due to the effect of the FDU on the surface N abundances. The variation (decrease) 
of C plays a much smaller role, but has the same qualitative effect of the increase of N, that is to shift the
RGB to larger values of $C_{F343N,F438W,F814W}$.

The dotted lines display the P1 RGBs that do 
not account for the effect of the FDU on the spectral energy distributions due to the change of the surface abundances.
At 13.5~Gyr the no-FDU RGB
is almost coincident with the FDU case, because at this metallicity and age its effect on the surface abundances is very small
also for the P1 composition.
For the 6 Gyr case, the no-FDU P1 RGB runs parallel to the P2 one, and diverges steadily from the calculations that include 
the FDU.

To assess quantitatively the impact of the FDU on the interpretation of M19 results,
we have considered P1-P2 pairs of RGBs with $\Delta {\rm [N/Fe]_{ini}}$=0.8, ages between 3 and 13.5~Gyr
for [Fe/H]=$-$1.3, and between 1 and 6 Gyr for [Fe/H]=$-$0.35.  
We have distributed randomly 100 stars with a 1:1 ratio between P1 and P2 objects, along the isochrones' RGBs
employing a Salpeter mass function --the choice of the mass function is not critical, because of the extremely narrow
mass range involved-- 
in the $M_{F438W}$ range of Fig.~\ref{figCMD}. We have then added to each synthetic star photometry a 
typical Gaussian photometric error taken from the artificial star tests, that is approximately the same for all
stars of the cluster sample, in this representative magnitude range.
The number of synthetic stars is roughly the same as the number of stars employed in M19 analysis.
For each pair of P1-P2 stars we have then calculated the $C_{F343N,F438W,F814W}$ distribution and determined
the 1$\sigma$ dispersion $\sigma(C_{F343N,F438W,F814W})^{RGB}$ --as in M19-- repeating the procedure 100 times to
determine its average value.

\begin{figure}
	\includegraphics[width=\columnwidth]{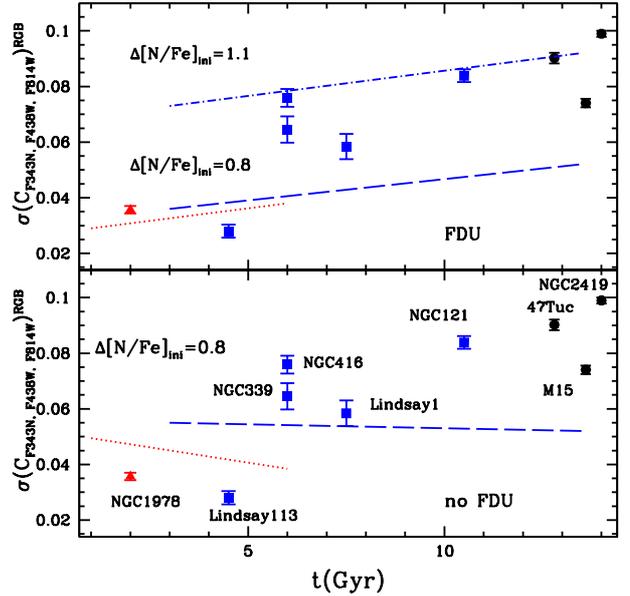}
        \caption{{\sl Top  panel:} Theoretical 
          $\sigma(C_{F343N,F438W,F814W})^{RGB}$ as a function of age for [Fe/H]=$-$0.35 (dotted line), 
          [Fe/H]=$-$1.3 (dashed line) and $\Delta {\rm [N/Fe]_{ini}}$=0.8.
          The dash-dotted line displays the results for
          $\Delta {\rm [N/Fe]_{ini}}$=1.1 and [Fe/H]=$-$1.3 (see text for details). 
          We also show
          the $\sigma(C_{F343N,F438W,F814W})^{RGB}$ values determined by M19 
          for a sample of MC clusters with photometric detection of MPs
          (filled red triangles for clusters with [Fe/H] around $-$0.35, and
          filled blue squares for clusters with [Fe/H] around $-$1.3).
          Results for three GCs are also displayed for comparison (filled black circles). 
          {\sl Bottom panel:} The same as the top panel but with theoretical $\sigma(C_{F343N,F438W,F814W})^{RGB}$
          values
          calculated for $\Delta {\rm [N/Fe]_{ini}}$=0.8 
          without accounting for the effect of the FDU on the spectral energy distributions.}
    \label{figsigmacunbi}
\end{figure}

The top panel of Fig.~\ref{figsigmacunbi} displays our theoretical $\sigma(C_{F343N,F438W,F814W})^{RGB}$ 
as a function of age, together with $\sigma(C_{F343N,F438W,F814W})^{RGB}$ measured by 
M19 for their sample of Magellanic Cloud clusters that show MPs, plus M19 determinations for
the Milky Way GCs 47~Tuc, M15 and NGC2419.
These bimodal MPs with 1:1 ratio and a constant $\Delta {\rm [N/Fe]_{ini}}$=0.8 display a trend with age
qualitatively similar to the observations. This is due to the effect of the FDU on the surface N abundances discussed above.
However, a constant $\Delta {\rm [N/Fe]_{ini}}$ with age does not match the observed average slope, for 
a linear fit to the observed $\sigma(C_{F343N,F438W,F814W})^{RGB}$ values 
gives a slope equal to 0.0060$\pm$0.0005~mag/Gyr, whilst the theoretical results displayed in the figure
have slopes equal to 0.0018~mag/Gyr
and 0.0015~mag/Gyr for [Fe/H]=$-$0.35 and [Fe/H]=$-$1.3, respectively. This implie
that the observed trends
are explained by a combination of both the effect of the FDU at constant $\Delta {\rm [N/Fe]_{ini}}$, and
a general increase of $\Delta {\rm [N/Fe]_{ini}}$ with age.
  
To clarify this point, the same figure displays also the $\sigma(C_{F343N,F438W,F814W})^{RGB}$ values for 
the same type of bimodal MPs (1:1 ratio
between P1 and P2 stars), but calculated considering P2 models
with $\Delta {\rm [N/Fe]_{ini}}$=1.1 (and [C/Fe]=[O/Fe]=$-$1.0, to preserve the CNO sum).
These values approximately match the upper envelope of the observed distribution,
with $\Delta {\rm [N/Fe]_{ini}}$=1.1 that roughly agrees with the range 
measured spectroscopically in 47~Tuc \citep{carretta05,marino16}. 

The theoretical $\sigma(C_{F343N,F438W,F814W})^{RGB}$ values at fixed age, [Fe/H] and $\Delta {\rm [N/Fe]_{ini}}$
depend of course on the statistical distribution of the [N/Fe] abundances within the prescribed range.
Still considering a bimodal distribution in terms of [N/Fe], when changing the P1/P2 ratio from
1:1 to a population with 70\% P1 stars and 30\% P2 stars with $\Delta {\rm [N/Fe]_{ini}}$ either 0.8 or 1.1~dex, the 
$\sigma(C_{F343N,F438W,F814W})^{RGB}$ dispersions are reduced by only a few 0.001~mag.
A larger reduction (by up to 0.03~mag in the case of $\Delta {\rm [N/Fe]_{ini}}$=1.1) 
is found when the [N/Fe] abundances are distributed uniformly within the prescribed $\Delta {\rm [N/Fe]_{ini}}$ range, a sort of
extreme opposite case compared to a 1:1 bimodal [N/Fe] distribution. However, even in this case the increase of
$\sigma(C_{F343N,F438W,F814W})^{RGB}$ with age is preserved.
Equal-weight multiple
subpopulations quantized in terms of [N/Fe] --depending on their number--  lead to
intermediate results between the bimodal 1:1 population, and the case of uniform [N/Fe] distribution.

The lower panel of Fig.~\ref{figsigmacunbi} displays the theoretical $\sigma(C_{F343N,F438W,F814W})^{RGB}$ values for 
the same bimodal MPs
calculated with $\Delta {\rm [N/Fe]_{ini}}$=0.8 and a 1:1 ratio between P1 and P2 stars, 
but without FDU. The absolute values and trends of $\sigma(C_{F343N,F438W,F814W})^{RGB}$ with age are different
from the \lq{correct \rq} ones that include the effect of the FDU.
The $\sigma(C_{F343N,F438W,F814W})^{RGB}$ values are higher than the FDU case at the younger ages,
showing a general anticorrelation with age. This of course impacts the determination of
$\Delta {\rm [N/Fe]_{ini}}$ from the measured $\sigma(C_{F343N,F438W,F814W})^{RGB}$ values in a sample of clusters.

\section{Conclusions}
\label{conclusions}

The impact of the FDU on the observed width of RGBs in old and intermediate-age clusters hosting MPs has been so far
largely unexplored. We have addressed this issue by considering several N-sensitive
pseudocolours employed to detect MPs in clusters of various ages.
In all cases, for a given initial difference $\Delta {\rm [N/Fe]_{ini}}$ between coeval P1 and P2 populations,
the effect of the FDU is to reduce the predicted RGB width compared to the case when the effect of the FDU on the 
spectral energy distribution is neglected.
The reduction is a function of age, and has also some dependence on [Fe/H]. These
effects stem from the dependence of the FDU
variations of the surface N and, to a lesser degree, C abundances,
on age and metallicity in models with P1 compositions. 

In the specific case of the pseudocolour employed by M19, when the FDU is accounted for, a constant $\Delta {\rm [N/Fe]_{ini}}$ produces
a general
increase with age of the predicted $\sigma(C_{F343N,F438W,F814W})^{RGB}$, qualitatively similar with what is observed, but 
with a shallower slope. The observed trend with cluster age 
can be matched only by a combination of both the effect of the FDU at constant $\Delta {\rm [N/Fe]_{ini}}$, and
a general increase of $\Delta {\rm [N/Fe]_{ini}}$ with age.
When the theoretical spectral energy distributions do not account for the FDU, 
the values of $\Delta {\rm [N/Fe]_{ini}}$ required to match the observed RGB widths will be smaller.
This effect only becomes significant for ages 
below $\sim$10~Gyr, when the FDU starts to alter more significantly the surface chemical composition--
see, e.g., the case of NGC1978 in Fig.~\ref{figsigmacunbi}.

The FDU affects also the quantitative interpretation, in terms N-abundance spreads, 
of the chromosome maps, when employed to identify MPs in intermediate-age
clusters. In the case of GCs, due to their old ages, the FDU is largely unable to affect appreciably the surface
N and C abundances, hence its effect on
the $C_{F275W, F336W, F438W}$, $C_{F336W, F343N, F438W}$, or $C_{F275W, F343N, F438W}$
pseudocolours is negligible. At younger ages the FDU
can affect the predicted values much more appreciably.
Hence, also in case of the chromosome maps, neglecting the FDU abundance changes can lead to an  
underestimate of the initial [N/Fe] spread for a given measured value of the RGB width.

The surface chemical changes due to the FDU play therefore an important role in the interpretation of the observed width of RGBs in
intermediate age clusters. It needs to be properly accounted for when determining initial N-abundance
spreads from photometry, and potential correlations with age or other cluster parameters.


%
%
%

\section*{Acknowledgements}
CU, NB, and SM gratefully acknowledge financial support from the European Research Council (ERC-CoG-646928, Multi-Pop).
NB also acknowledges financial support from the Royal Society (University Research Fellowship). 
Support for ICZ was provided by NASA through Hubble Fellowship grant HST-HF2-51387.001-A awarded by the Space Telescope
Science Institute, which is operated by the Association of Universities for Research in Astronomy, Inc., for NASA,
under contract NAS5-26555.
SC acknowledges support from Premiale INAF MITiC, from INFN (Iniziativa specifica TAsP), and grant AYA2013-
42781P from the Ministry of Economy and Competitiveness of Spain. 




\bibliographystyle{mnras}








\bsp	
\label{lastpage}
\end{document}